\documentclass[twocolumn]{revtex4-1}
\usepackage{graphicx, dblfloatfix}
\usepackage[latin1]{inputenc}
\usepackage{mathtools}
\usepackage{amsmath}
\usepackage{amsfonts}
\usepackage{upgreek}
\usepackage{color}
\usepackage{subfigure}
\renewcommand{\figurename}{Fig.}

\begin{document}

\title{Unconventional magnetization textures and domain-wall pinning in Sm--Co magnets}

\author{L. Pierobon\textsuperscript{1}, A. Kov\'acs\textsuperscript{2}, R. E. Sch\"aublin\textsuperscript{1,4}, S. S. A. Gerstl\textsuperscript{1,4}, J. Caron\textsuperscript{2}, U. Wyss\textsuperscript{3}, R. E. Dunin-Borkowski\textsuperscript{2}, J. F. L\"offler\textsuperscript{1}, M. Charilaou\textsuperscript{1,5}}


\affiliation{\textsuperscript{1}Laboratory of Metal Physics and Technology, Department of Materials, ETH Zurich, 8093 Zurich, Switzerland}
\affiliation{\textsuperscript{2}Ernst-Ruska Centre for Microscopy and Spectroscopy with Electrons and Peter Gr\"unberg Institute, Forschungszentrum J\"ulich, 52425 J\"ulich, Germany}
\affiliation{\textsuperscript{3}Arnold Magnetic Technologies, 5242 Birr-Lupfig, Switzerland}
\affiliation{\textsuperscript{4}Scientific Center for Optical and Electron Microscopy, ETH Zurich, 8093 Zurich, Switzerland}
\affiliation{\textsuperscript{5}Present address: Department of Physics, University of Louisiana at Lafayette, Lafayette, 70504 LA, USA}


\begin{abstract}
The most powerful magnets for high temperature applications are Sm--Co-based alloys with a microstructure that combines magnetically soft and hard regions. The microstructure consists of a dense domain-wall-pinning network that endows the material with remarkable magnetic hardness. A precise understanding of the coupling between magnetism and microstructure is essential for enhancing the performance of Sm--Co magnets, but experiments and theory have not yet converged to a unified model. Here, we combine transmission electron microscopy, atom probe tomography, and nanometer-resolution off-axis electron holography with micromagnetic simulations to show that the magnetization processes in Sm--Co magnets result from an interplay between curling instabilities and pinning effects at the intersections between magnetically soft and hard regions. We also find that topologically non-trivial magnetic domains separated by a complex network of domain walls play a key role in the magnetic state. Our findings reveal a previously hidden aspect of magnetism and provide insight into the full potential of high-performance magnetic materials. \end{abstract}

\maketitle


\section{Introduction}
Sm--Co-based materials are the strongest magnets available today, particularly for vital high-temperature precision applications thanks to their high Curie temperatures \cite{hadjipanayis2000,gutfleisch2011,mccallum2014}. Extensive research and industrial development in the past few decades has led to a significant improvement in their magnetic performance \cite{Horiuchi2013}. One such example is a highly engineered Sm--Co-based system that consists of a cellular microstructure with a Sm$_2$Co$_{17}$ matrix that is enclosed by SmCo$_5$ cells and intersected by the so-called Z phase (Zr-rich platelets) perpendicular to the $c$-axis of the Sm$_2$Co$_{17}$ matrix, which is at the same time the magnetic easy axis \cite{hadjipanayis1999,hadjipanayis2000,duerrschnabel2017}. This characteristic geometry results from tailored aging-heat treatments \cite{hadjipanayis2000}, and corresponds to a network of intertwined magnetically soft (Sm$_2$Co$_{17}$) and hard (SmCo$_5$) regions.

Magnetic properties due to combinations of hard and soft structures are highly tunable because they make use of the high saturation magnetization of the soft phase and the strong magnetocrystalline anisotropy of the hard phase \cite{kronmueller1996,Skomski2013}. Particular attention has been devoted to modeling the Sm--Co cellular microstructure \cite{skomski1997,Fidler2000,streibl2000,Fidler2004} in order to predict its coercivity. The enhanced coercivity in these cellular Sm--Co magnets emerges from the difference between the magnetocrystalline anisotropy of the two phases and consequently the difference in domain-wall energy \cite{skomski1997}. While conventional wisdom states that this microstructure constitutes a pinning system for domain walls, the exact magnetization processes remain elusive despite the intense activities that have been performed to understand the interaction of domain walls with the SmCo$_5$ cells \cite{Gaunt1972,livingston1977,Nagel1979,hadjipanayis1982b,fidler1982,wong1997,Zhang2018}. 

Magnetic imaging experiments, by means of Lorentz transmission electron microscopy (LTEM), magnetic force microscopy, and Kerr microscopy, have revealed that domain walls follow the SmCo$_5$ cell morphology \cite{fidler1982,hadjipanayis1982,hadjipanayis2000b,Gutfleisch2006,okabe2006,SepehriAmin2017,Zhang2018}, thus confirming strong pinning at the cell boundaries. Theory and experiment, however, have yet to converge on the role of the Z phase on the magnetic properties of this material \cite{skomski1997,Fidler2000,streibl2000,Fidler2004}. Forward modeling of the Z phase is impeded by the fact that the material parameters of this phase cannot be easily estimated because the thickness of the Z phase can be as thin as 1-2 atomic layers (the thickness varies from material to material), and thus cannot be compared with measurements on bulk samples \cite{katter1996}. Hence, high-resolution imaging of the magnetization textures is crucial to elucidate the interplay between the soft, hard, and Z phases, and unveil the magnetization processes that are at play in the cellular Sm--Co magnets.

Here we present a detailed study of the magnetic state in a cellular Sm--Co magnet containing Fe, Cu and Zr, where we correlate atomic-resolution TEM and atom probe tomography (APT) with high-resolution LTEM and off-axis electron-holography (EH) imaging of the domain-wall structure, and systematically compare it with detailed micromagnetic simulations. By matching experiments and theory one-to-one we show that the nanoscale magnetization processes in cellular Sm--Co magnets stem from an interplay between pinning at the SmCo$_{5}$ cell boundaries and curling instabilities at the intersections between all three phases.

\section{Results and Discussion}

The Sm--Co sample in our study has an overall chemical composition of Sm(Co,Fe,Cu,Zr)$_{7.57}$ with minor amounts of oxygen (see the Methods section). Figure 1a shows for this sample an overview of the microstructure, revealing the typical soft-magnetic Sm$_2$Co$_{17}$ matrix enclosed by the hard-magnetic SmCo$_5$ cells and intersected by the Z-phase platelets. A close-up view in Fig. 1b and corresponding energy-dispersive X-ray spectroscopy (EDX) chemical maps in Fig. 1c--g show interfaces between the three phases and confirm that the Z-phase is rich in Zr. The thickness of the interface between the Sm$_2$Co$_{17}$ matrix and the SmCo$_5$ cells ranges from atomically sharp to 2 nm (visible as blurry contrast), whereas the interfaces with the Z phase are always atomically sharp. The $c$-axis of the crystal structure (see Fig. 1 in the Supplementary Material) lies inside the TEM lamella, and the Z platelets are always perpendicular to the $c$-axis \cite{duerrschnabel2017}.

The APT reconstruction in Figure 1h shows the isoconcentration surfaces of Zr and Fe with concentration values of 9.8 and 13.5 at. \%, respectively, and reveal four perfectly flat Z platelets, where two rightmost of which are actually so close that they are visible only as one wide platelet. In the top middle part of Fig. 1h between two Z platelets a twisted SmCo$_5$ cell can be seen. The twisted shape explains why in Fig. 1a different interfaces between the 2:17 and 1:5 phases have different sharpness. The concentration profiles of individual elements across a SmCo$_5$ cell and a Z platelet are shown in Figures 1i and j, respectively (across the blue areas in the inset figures). Cu is found inside the cell with a Gaussian-like distribution, which might critically affect the magnetic properties \cite{SepehriAmin2017}. The increase in Sm across the SmCo$_5$ cell appears to be non-symmetrical, which may be due to applying a one-dimensional concentration profile to the twisting of the cell. As expected, Zr increases across the Z platelet, but, surprisingly, Cu segregates at the interfaces between the platelet and the matrix, which has a significant impact on the magnetic performance. Therefore, further nanoscale segregation and clustering studies are feasible and encouraged. The SmCo$_5$ cells are typically around 200 nm wide across their widest region and approximately 15 nm thick (Fig. 1a), and the Z platelets are at most 5 nm thick (Fig. 1b) and thus consist of only a few atomic layers (some platelets consist of only 1--2 atomic lattice planes). As we will discuss below, it is the thickness of the Z platelets that plays a crucial role in the magnetic properties.

\onecolumngrid

\begin{figure}[b]
\centering
\includegraphics[width=0.8\columnwidth]{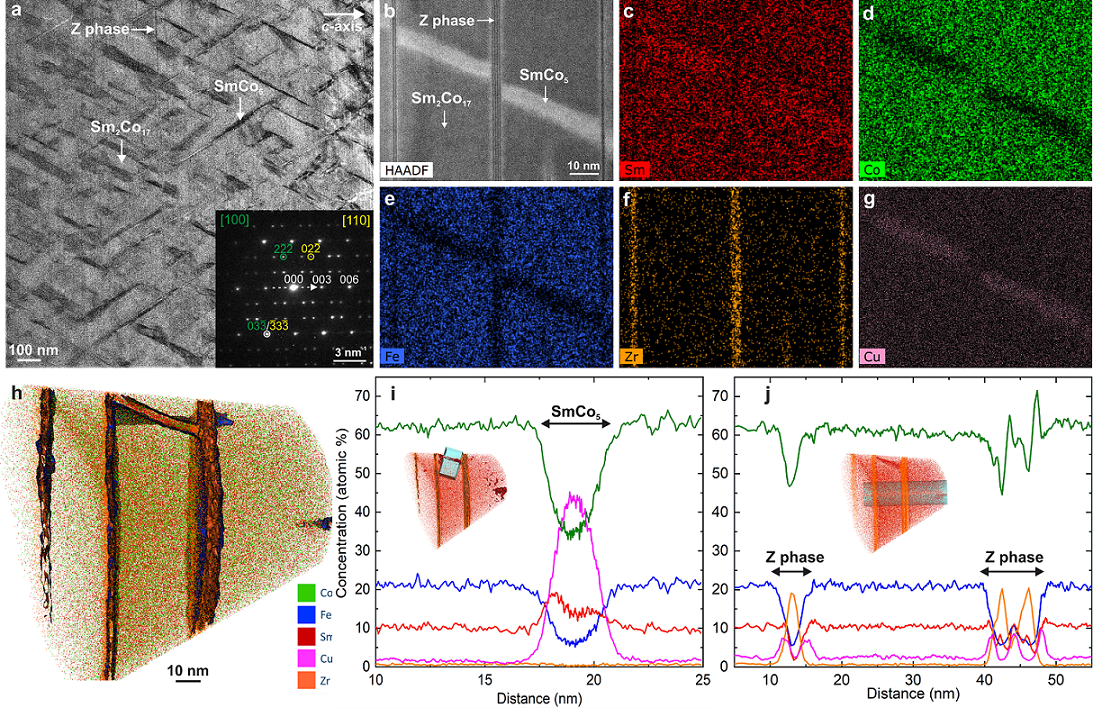}
\caption{\textbf{Sm--Co microstructure.} \textbf{a} Bright-field TEM image of a Sm--Co magnet showing the Sm$_2$Co$_{17}$ matrix (light grey) enclosed by SmCo$_5$ cells (dark grey), with the entire structure intersected by the Z phase. The corresponding diffraction pattern, shown in the bottom right corner, contains reflections from the [100] (green) and [110] (yellow) directions (see Suppl. Fig. 1). \textbf{b} High-angle annular dark-field (HAADF) scanning TEM image showing the details of the microstructure, accompanied by EDX chemical maps of \textbf{c} Sm, \textbf{d} Co, \textbf{e} Fe, \textbf{f} Zr and \textbf{g} Cu. \textbf{h} Atomic-resolution APT reconstruction with the isoconcentration surfaces of Zr and Fe, exhibiting flat Z-phase platelets (vertical) and a twisted SmCo$_5$ cell (in the top middle between the two flat Z phase platelets). Concentration plots, as indicated by insets, of individual elements show that \textbf{i} Cu accumulates in the middle of the cell, while Sm increases non-symmetrically across the 1:5 cell, and \textbf{j} Zr peaks in the middle of the the Z phase, while Cu accumulates at the interface between the Z phase and the Sm$_2$Co$_{17}$ matrix.}
\label{stem}
\end{figure}
\newpage

\begin{figure}
\centering
\includegraphics[width=0.96\columnwidth]{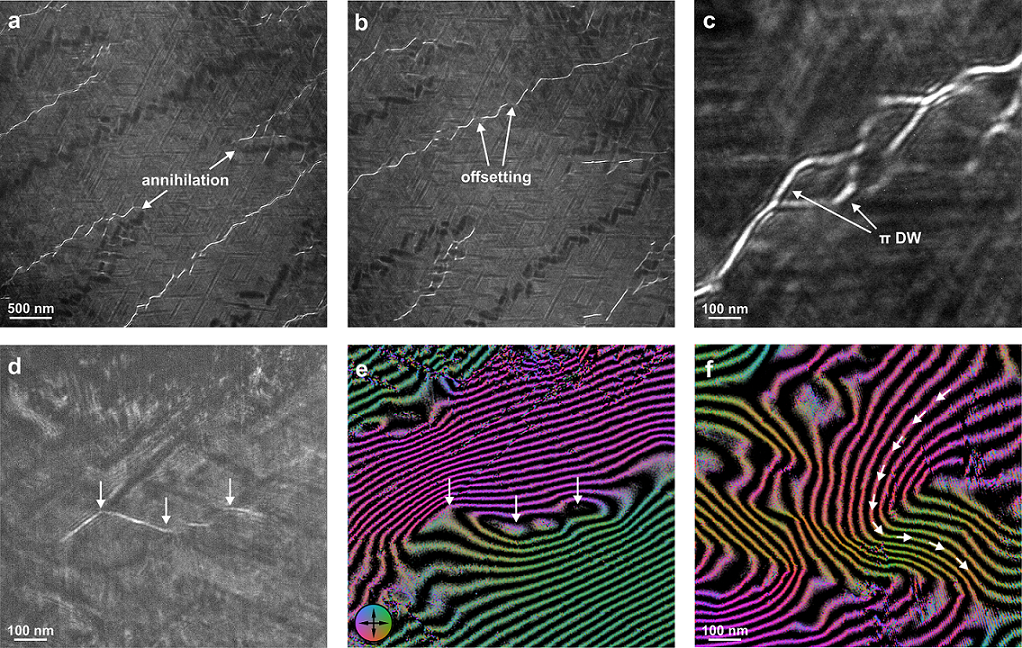}
\caption{\textbf{Magnetic structures in a Sm--Co magnet.} DW structures imaged by LTEM in 0.5 mm \textbf{a} over- and \textbf{b} underfocus are characterized by domain-wall annihilation and offsetting. \textbf{c} Higher magnification reveals a topologically non-trivial structure with branching and alternating $\uppi$ and $2\uppi$ magnetization profiles. A comparison of \textbf{d} a Fresnel defocus image and \textbf{e} a magnetic induction map extracted from off-axis EH of the same region (arrows indicate the same location) shows a $\uppi$ domain wall involving vortex-like curling. \textbf{f} Magnetic induction map of a complex magnetic state consisting of DWs with a wide range of angles (the arrows follow the magnetic induction direction). The phase difference between adjacent contours in the induction maps is 2$\uppi$ radians, and the different contour spacing in panels e and f results from different specimen thickness.}
\label{lorentz}
\end{figure}
\twocolumngrid

Figure 2 shows magnetic imaging studies of the cellular Sm--Co magnet in a thermally demagnetized (magnetically pristine) state. Figures 2a and b are over- and underfocus Fresnel-mode LTEM images of the same area, respectively. Domain walls (DWs) appear as alternating sharp and bright contrast (convergent), or blurry and dark contrast (divergent), depending on the sign of the defocus and the sense of the magnetization profile along the domain wall. Changing the sign of the defocus inverts the contrast, i.e. the same domain walls have opposite contrast in over and underfocus images. These LTEM images reveal a complex DW pattern that follows exactly the microstructure of the material, similar to those shown in the literature \cite{hadjipanayis1982,fidler1982,hadjipanayis2000b,Zhang2018}. Specifically, we observe that the domain walls are always adjacent to the SmCo$_5$ cells and are offset by approximately 50 nm via the Z phase exactly where they intersect the cells. Even though LTEM does not enable a quantification of the magnetization direction along the domain-wall profile, considering the high uniaxial anisotropy in the system we may assume that these are $\uppi$ Bloch-type domain walls \cite{fidler1982}. The average domain-wall thickness width was estimated from a series of images taken at different defocus values (see Fig. 2 in the Supplementary Material) to be $4\pm 2$ nm. This is in agreement with theory, considering the properties of Sm$_2$Co$_{17}$ from which $\delta_\mathrm{dw} = \sqrt{A/K_\mathrm{u}} = 2.7$ nm can be derived, where $A$ is the exchange stiffness and $K_\mathrm{u}$ is the uniaxial magnetocrystalline anisotropy (see Methods). This result suggests that the domain walls are mostly located in the soft Sm$_2$Co$_{17}$ phase, reminiscent of an exchange spring magnet \cite{fullerton1999}.

While a zig-zag domain-wall structure is well known {\cite{Tian2015}, here we have observed for the first time unexpected domain wall patterns at some of the intersections between the three phases, where domain walls with opposite sense meet and annihilate each other, leaving a trivial ferromagnetic state (marked with arrows in Figs. 2a and b). Surprisingly, topologically complex structures bounded with two $\uppi$ walls of the same sense, i.e. total winding of 2$\uppi$ can also be observed, as shown in Fig. 2c. The unwinding of such regions is non-trivial and requires a violation of topological constraints. 

\newpage

We have complemented LTEM with off-axis EH to gain in-depth information on the direction of the local magnetic field inside the sample. Figure 2d shows an LTEM image of a domain wall pinned to a SmCo$_5$ cell with contrast that varies in intensity. Domain walls may be tilted inside the sample and therefore overlap with adjacent magnetic domains, which might result in such contrast. However, a magnetic induction map extracted from off-axis EH of the same area shown in Fig. 2e gives more information on this magnetic structure. The domain wall has a winding of $\uppi$ and the magnetic field curls around the SmCo$_5$ cell, forming closed loops reminiscent of magnetic vortices (indicated by the middle arrow). This curling may explain the variation of contrast intensity in LTEM. Interestingly, Figure 2e reveals another domain wall in the top left corner, which is not visible in LTEM in Fig. 2d. Figure 2f shows that the magnetic texture in some areas can be so complex that some domain walls do not have a well-defined angle; instead the angle varies between $\uppi/2$ and $\uppi$. These exotic magnetization textures are closely correlated with the microstructure and indicate that topological aspects need to be considered in order to correctly interpret the magnetic state.

Given the elaborate microstructure, these observations raise various questions regarding the magnetic state in cellular Sm--Co magnets. In order to obtain further insight, we performed detailed high-resolution micromagnetic simulations to elucidate the formation of the observed complex domain patterns. To this end, it is imperative that we fully consider the real microstructure, and thus we constructed a simulation system directly from the TEM images shown above. Figures 3a,b show how we have truncated the microstructure in order to model a system with the three phases at the same geometry and scale.

In our simulations, we have considered the ferromagnetic exchange and uniaxial anisotropy energies, the dipole-dipole interactions, and the exchange energy between the three different phases. The material parameters ($A$, $K_\mathrm{u}$, and saturation magnetization $M_\mathrm{S}$) are well known and were taken from the literature \cite{katter1996,nagel2003,SepehriAmin2017,duerrschnabel2017} (see Methods). The exchange interaction energy between the three phases is unknown. We performed parametric micromagnetic studies where we varied the exchange interaction and compared the theoretical hysteresis curve with the experimental data, thus deducing the correct values by matching simulations to experiments (Fig. 3c-e).

As mentioned above, the precise material properties of the Z phase are unknown because the exact chemical composition is unclear and the platelets can be as thin as a single atomic layer (see Fig. 1b). All material parameters, e.g., exchange stiffness, saturation magnetization, and magnetocrystalline anisotropy are affected by the reduced dimensions of the platelets and their interfaces \cite{Hellman2017}. It is known from thin-film studies that the anisotropy is the most sensitive property and changes drastically depending on the thickness and local atomic arrangements \cite{Charilaou2016}. We have therefore performed simulations where we considered the Z phase to be either anisotropic or isotropic. In the former case, we assigned the bulk value of $K_\mathrm{u}$ \cite{katter1996,duerrschnabel2017}, whereas in the latter case we set $K_\mathrm{u}=0$. From the comparison of experimental and theoretical $M(H)$ curves, we find that the simulations match the experimental observations only if $K_\mathrm{u}=0$. Hence, for the rest of the discussion we assume that $K_\mathrm{u}=0$ for the Z phase.

In order to make our simulation quantitatively comparable to the experimental results, we matched the theoretically predicted coercivity with the experimental one by varying the exchange energy between the three phases. We find that the coercivity increases with decreasing exchange energy (see Fig. 3c). This supports experiments that show that increasing Cu content leads to higher coercivity, depending on the compositional gradient at the boundary \cite{Gutfleisch2006,SepehriAmin2017}. This is due to the formation of Cu-rich interfaces between the magnetic phases that decrease the exchange coupling.

We have also found that changing the thickness of the SmCo$_5$ cells does not modify the magnetic performance strongly, in agreement with experiments \cite{Shen2016} and theory \cite{Nishida2012}, showing that the pinning field is saturated for a SmCo$_5$ thickness of more than 4 nm. This contradicts the predictions by Fidler et al. \cite{Fidler2000,Fidler2004}, stating that the SmCo$_5$ thickness should be at least three times the exchange length ($3\delta_\mathrm{exc} \approx 20$ nm) for effective domain-wall pinning. The pinning, however, is a complex process and depends strongly on the Cu content in the SmCo$_5$ cells \cite{Kronmueller2002}. In our experiments we have found a Cu-composition gradient, but because the variation of the magnetic material parameters as a function of Cu is unknown, we modeled SmCo$_5$ with a homogeneous Cu enrichment.

Furthermore, we have found that the coercivity strongly depends on the Z-phase thickness (see Fig. 3d). In the absence of the Z phase, it has a maximum value of 5.7 T, and decreases significantly with increasing Z-phase thickness up to 5 nm, where it reaches a minimum of 3 T and then remains constant. These results explain recent experimental observations, where the magnetic performance deteriorated with increasing Z-phase thickness \cite{Zhang2018}. However, since Zr is essential in forming the Sm--Co microstructure \cite{hadjipanayis1982}, it cannot be completely eliminated from the material, but the Z-phase platelets may be designed to be as thin as possible to maximize the performance. As we will discuss below, smaller thickness impedes magnetization curling and hence a stronger external field is required to initiate the magnetization reversal process, which begins at the intersections of the Z and Sm$_2$Co$_{17}$ phases.

\onecolumngrid
\newpage

\begin{figure}
\centering
\includegraphics[width=1.0\columnwidth]{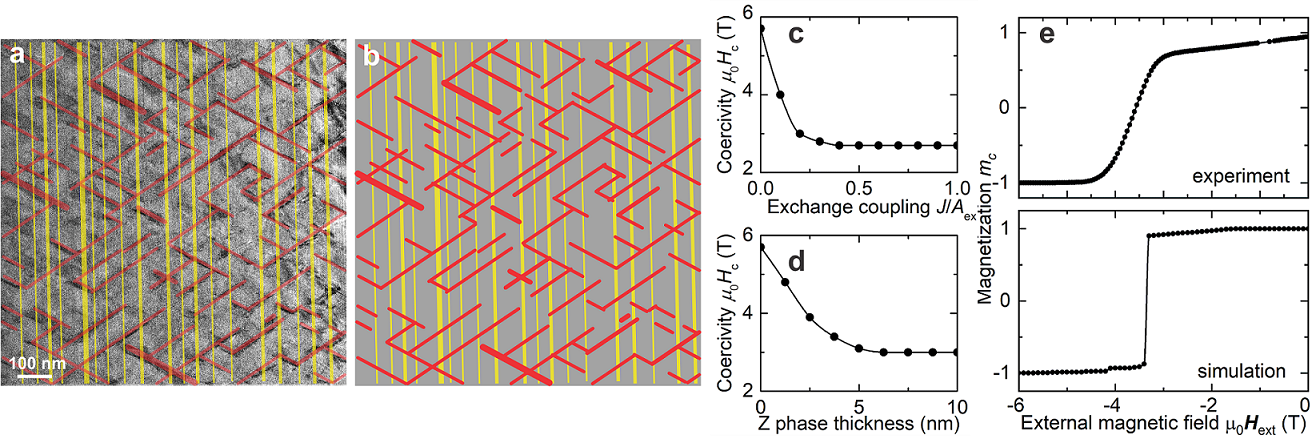}
\caption{\textbf{Modelling the microstructure of Sm--Co magnets.} \textbf{a} Modelling of the microstructural features in Sm--Co, as seen in Fig. 1a, to create \textbf{b} a model with the Sm$_2$Co$_{17}$ matrix (grey), the SmCo$_5$ cells (red), and the Z phase (yellow). \textbf{c} Simulated dependence of the coercivity as a function of exchange coupling between the individual phases, illustrating that smaller exchange between the phases leads to higher coercivity. Using these results, we can compare the theoretical coercivity with that of real samples. \textbf{d} Simulated dependence of the coercivity on the thickness of the Z phase, showing that it increases significantly with decreasing thickness. \textbf{e} Comparison between an experimentally measured $M(H)$ loop at $T=300$ K and a simulated loop along the easy axis. By matching the simulation to the 300 K experimental data, we have obtained the value of the exchange stiffness in the system. The external field is applied parallel to the $c$ axis, i.e., perpendicular to the Z-phase platelets. }
\label{model}
\end{figure}
\twocolumngrid

In the simulations that we discuss in the following section, the thicknesses of the cell boundary and the Z phase were derived directly from the TEM images, where we have 10 nm thick SmCo$_5$ cells and 1 to 5 nm thick Z-phase platelets. Figure 3e shows an experimentally measured (see Methods) and a simulated $M(H)$ curve along the easy axis, confirming the agreement between experiment and theory, specifically the value of the remanence $M_\mathrm{R} = 0.95 M_\mathrm{S}$ and a gradual decrease of the magnetization upon reversing the external field prior to the full magnetization switching at 3.4 T. Our simulations indicate that the gradual decrease is due to the magnetically isotropic defects that reverse their magnetization earlier than the rest of the material, which is yet another confirmation that the Z phase does not exhibit significant magnetocrystalline anisotropy. Note that the $M(H)$ curves are not identical, because the experiment was performed on a bulk sample, where we have gradual switching of parts of the material, while the theory considers a single thin lamella. Furthermore, the lamella with a thickness/width aspect ratio of about 1/1000 has an additional shape anisotropy, though much smaller than the magnetocrystalline anisotropy $K_\mathrm{u}>>\mu_0 M_\mathrm{s}^2/2$.

In order to study the magnetization texture in cellular Sm--Co magnets, we compare simulated domain-wall structures with those observed in the experiment. Fig. 4a shows an experimental Fresnel defocus image at 0.24 mm overfocus of the region shown in Fig. 3 in the thermally demagnetized state, which contains magnetic domains separated by three domain walls with the characteristic zig-zag shape following the microstructure. In our simulations, we initialized the system with three straight DWs and ran it for 1 ns to allow the domain walls to relax into the state of minimal energy. The resulting relaxed magnetization texture is overlaid in Fig. 4b onto its corresponding TEM image. The positions of the domain walls in the experimental LTEM image (white intersected lines) and the simulated magnetization image match very closely. In fact, in both cases the domain walls follow precisely the microstructure. Additionally, small magnetic domains with opposite magnetization approximately 5 nm wide are present (white circles in Fig. 4), which form due to strong pinning to SmCo$_5$. In order to directly correlate the micromagnetics with experimental observation, a magnetic phase shift image has been calculated based on the micromagnetic results (see Methods). From it, a Fresnel image and a magnetic induction map were reconstructed and shown in Figs. 4c and d, respectively. A close match between the theory and experiment is apparent, as the simulated images contain features, such as domain wall offsetting and curling, identical to those observed in Fig. 2. Additionally, the micromagnetic simulation reveals that the curling is out-of plane (see Figure 5).

\onecolumngrid
\newpage
\begin{figure}
\centering
\includegraphics[width=1\columnwidth]{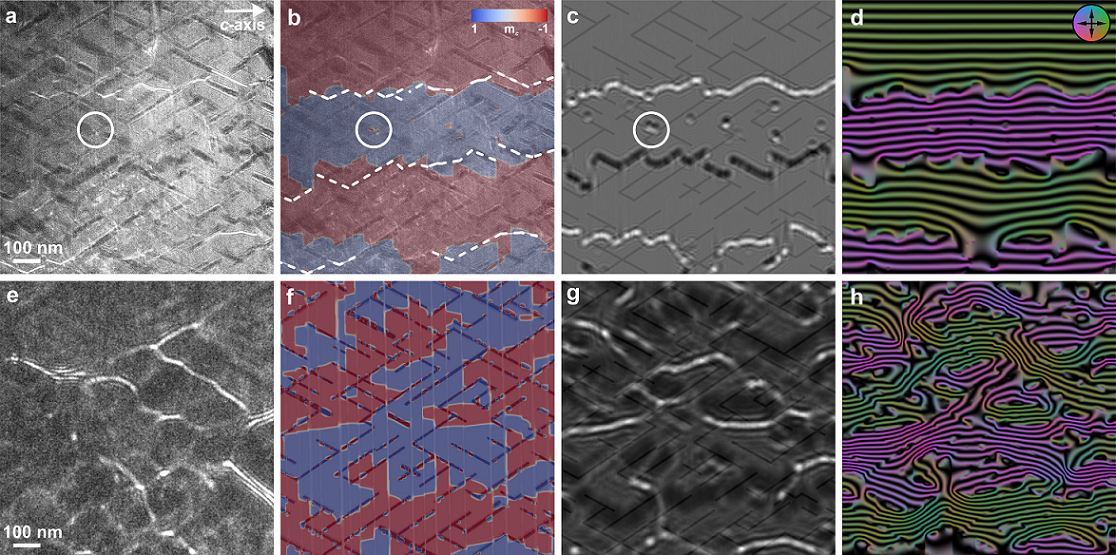}
\caption{\textbf{Comparison between experimental observations and theoretical predictions of the magnetization in Sm--Co.} \textbf{a} LTEM image at 0.24 mm overfocus of the region shown in Fig. 1a reveals four magnetic domains separated by DWs. \textbf{b} Micromagnetic simulation of magnetization in the same microstructure superimposed on Fig. 4a, and \textbf{c} a Fresnel defocus image and \textbf{d} magnetic-induction map simulated based on Fig. 4b show a distinct resemblance of the DW network in theory and experiment, namely the pinning at SmCo$_5$ and offsetting by the Z phase. The white circles indicate one of the small domains with opposite magnetization. \textbf{e} Fresnel defocus image at 0.8 mm overfocus of the remanent state (not the same region as panel a). \textbf{f} Simulation of the remanent state of panel a. Consecutive domains of opposite magnetization are present at the SmCo$_5$ cells. The corresponding \textbf{g} Fresnel defocus image and \textbf{h} magnetic induction map of the remanent state, once again illustrating very good agreement between experiment and theory, specifically the branching DW network pinned at SmCo$_5$. The phase difference between adjacent contours in the induction maps is 2$\uppi$.}
\label{Comparison}
\end{figure}
\twocolumngrid

Furthermore, we ramped up the magnetic field to saturate the sample in both the experiment and the simulation and then removed the field to observe the remanent state. As we know from Fig. 3e, this corresponds to $M = 0.95 M_\mathrm{S}$, meaning that one would expect the magnetization to be nearly uniform in the remanent state, but surprisingly this is not the case. Figure 4e, which shows an experimental Fresnel image of the remanent state (after applying an external field of 6 T) at 0.8 mm overfocus, reveals a state with a complex network of domains separated by branching domain walls (note that the region shown in Fig. 4e is not the same as in Fig. 4a.). This state represents initial nucleation stages of the magnetization process. Some domain walls have strong contrast with multiple lines of three or more satellites, which are only visible for $\uppi$ domain walls that are perfectly edge-on. Lower-degree domain walls usually form weak and fading satellites. In Fig. 4f, the simulated magnetization state, again very similar to the experiment, contains a large number of small domains with opposite magnetization pinned to the SmCo$_5$ cells. These are the smallest possible domains, around 5 nm wide, constrained by the domain-wall width. We have again simulated a magnetic phase image from the micromagnetic simulation, from which we extracted a Fresnel defocus image and a magnetic induction map, shown in Fig. 4g and h, respectively. The magnetic texture shows a good match with the experiment, notably the complex DW network. The magnetic induction map reveals the presence of vortex-like out-of-plane curling in the remanent state.

In the following, we take a deeper look into our simulations beyond the experimental limitations. Figure 5 shows the magnetization as contour plots overlaid onto the microstructure. Note the prominent resemblance of Fig. 5a with the Fresnel defocus images of Fig. 2, i.e. the DWs follow exactly the SmCo$_5$ cell geometry, including the offsetting by the Z phase platelets. Figure 5b shows a close-up image of the magnetization texture around intersections between the three phases for the region marked with a square in Fig. 5a. By fitting a magnetization profile across the domain wall with $\tanh\left(r/\lambda \right)$, where $r$ is the distance from the domain-wall center and $\lambda=\delta_\mathrm{dw}/2$, we have deduced the DW width in the SmCo$_5$ and the Sm$_2$Co$_{17}$ phases to be 1.5 nm and 4.7 nm, respectively. These values are slightly larger than the theoretically expected values of 1.2 nm and 2.7 nm using the equation $\delta_\mathrm{dw} = \sqrt{A/K_\mathrm{u}}$, but they agree with our experimental observation of the domain-wall thickness of 4 $\pm$ 2 nm (see Suppl. Fig. 2). This confirms our conclusion that the domain walls are mostly located in Sm$_2$Co$_{17}$ and pinned to the the SmCo$_5$ cell. The minimum domain size of 5 nm is also shown in Fig. 5b at the left edge of the SmCo$_5$ cell that is intersected by a Z phase platelet. Notably, the DWs inside the Z phase are extremely thin, and the moments turn away from the $c$-axis due to the dominating shape anisotropy of the platelets. This indicates that the DWs between the Sm--Co phases and the Z phase are in fact $\uppi/2$ DWs. To minimize the associated exchange energy, the magnetic moments are twisted with opposite handedness at the edges of the SmCo$_5$ cell boundary. This is further analyzed in Fig. 5c, which shows a detailed view of a region where the three phases intersect and a domain wall propagates through all of them. We observe a narrow domain wall in the SmCo$_5$ cell, a broader wall in the Sm$_2$Co$_{17}$ cell, and a curling of the moments away from the $c$-axis inside the Z phase. The curling has a significant out-of-plane component. The domain wall is in fact injected into the hard phase at the location where the three phases meet. These results shed new light on previous observations based on electron microscopy \cite{okabe2006}, which suggested an out-of-plane tilting of the magnetic flux away from the easy axis around intersections.

The points in the microstructure where the three phases intersect play a critical role in the magnetization process because their edges with different material properties enable curling instabilities. We show in Fig. 5e-h the process of magnetization reversal, i.e., coming from a saturated state and applying an external field in the opposite direction. The demagnetization starts at the intersections between the Z phase and the soft Sm$_2$Co$_{17}$ matrix in the form of nucleating domains that gradually grow inside the Sm$_2$Co$_{17}$ matrix, and become pinned by the hard SmCo$_{5}$ cells. The domain growth then progresses through the intersections where all three phases meet and crosses the SmCo$_{5}$ cells through these points. This further indicates that the Z phase does indeed play a vital role in the magnetization process, emerging as an interplay between curling instabilities at intersections between the Z phase and the Sm$_2$Co$_{17}$ matrix, and pinning at the hard cells. Importantly, this demagnetization process might also be responsible for the formation of DWs with higher winding angles, such as those observed in Fig. 2. 

\onecolumngrid

\begin{figure}[!hb]
\centering
\includegraphics[width=1.0\columnwidth]{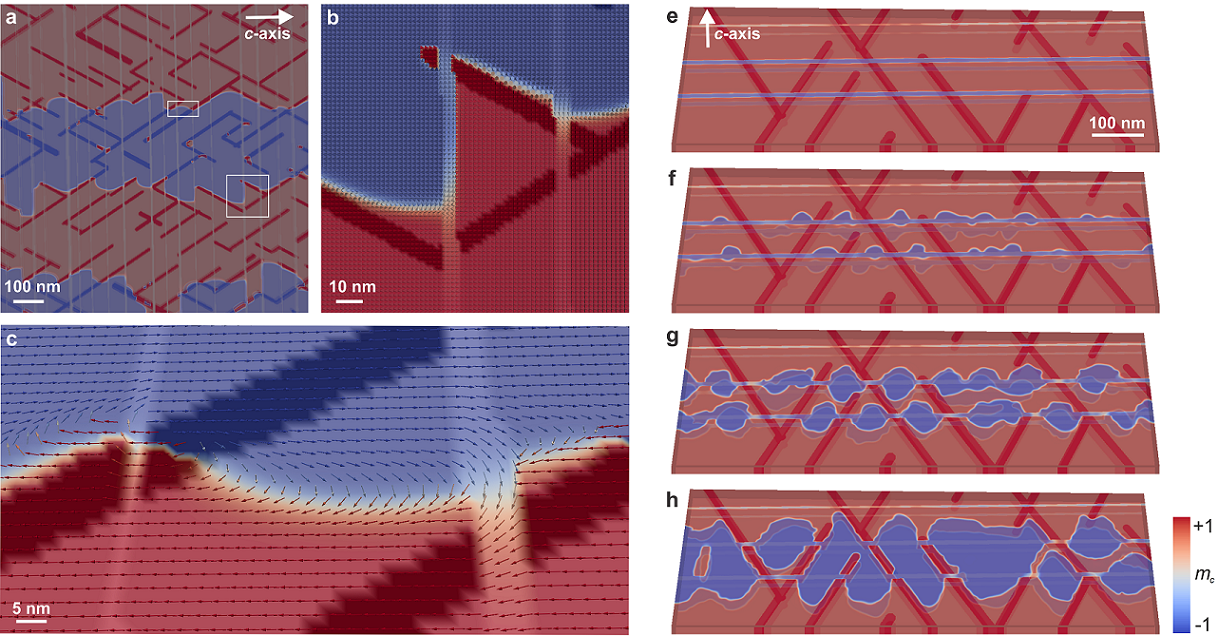}
\caption{\textbf{Simulations of 3D DW structures and their nucleation in Sm--Co magnets.} \textbf{a} Structure with four domains in a demagnetized state showing that DWs are pinned by the microstructural features, specifically the SmCo$_5$ cells, and are offset by the Z phase. \textbf{b} Close-up of the marked square area in \textbf{a} reveals that the domain walls nucleate at the phase boundary between Sm$_2$Co$_{17}$ and SmCo$_5$ and are mostly situated inside Sm$_2$Co$_{17}$. \textbf{c} Detailed view of the magnetization texture at the intersection of all three phases, showing how a DW can be injected into the hard phase through a Z-phase platelet, and consequently how its thickness varies between the two phases. \textbf{e--h} Illustration of the magnetization reversal process at a slightly supercritical field, i.e., larger than the switching field, parallel to the $c$-axis: as time progresses domains with magnetization parallel to the field begin nucleating at the Z phase and spread into the Sm$_2$Co$_{17}$ matrix, but their growth is impeded by the SmCo$_5$ cells (the time step between each figure is 0.1 ns).}
\label{simulations}
\end{figure}

\newpage
\twocolumngrid

\section{Conclusions}

We have shown via magnetic imaging in TEM, APT and micromagnetic simulations that there are sharp magnetic DWs in cellular Sm--Co magnets that follow exactly the morphology of the hard-phase cells and are offset by the Z-phase platelets. Based on our finding that the domain walls are mostly located in the soft Sm$_2$Co$_{17}$ phase, reminiscent of an exchange spring system \cite{kneller1991}, we propose that the thickness of SmCo$_5$ cells should be reduced to a minimum, considering that the domain walls in SmCo$_5$ are only about 1.5 nm wide. Importantly, we have confirmed that the Z phase plays a crucial role in the magnetization process because curling instabilities at the intersections between the soft and Z phase act as nucleation sites for DWs upon switching the external magnetic field. The domain walls propagate inside the soft Sm$_2$Co$_{17}$ matrix and become pinned at the SmCo$_{5}$ hard cells. In that case they can only propagate further through the intersections between the cells and the Z phase. Hence, we propose that both the Zr and Sm contents should be reduced in this system. This would result in finer Zr platelets, which would impede magnetization switching, and the reduced Sm would lead to smaller SmCo$_5$ cells, which would in turn increase remanence, leading to a much more powerful magnet. Additionally, we have observed topologically non-trivial domains with highly-complex DWs, and, where all three phases meet, out-of-plane curling of domain walls. These exotic magnetic structures need to be studied further in order to understand the physics of multi-phase magnets in detail and to allow fully harnessing the potential of these high-performance materials.

\section{Methods}

\subsection{Sample synthesis:}

For the synthesis of the Sm--Co magnet, the alloying elements were melted in an induction furnace under argon and the resulting alloy was cast in a metallic mold. After crushing the alloy with a hammer mill, the resulting powder was milled in a jet mill with a particle size of 4 -- 8 $\upmu$m. The powder was then filled into a rubber mold, aligned with magnetic pulses of field strength 5 T and pressed in an isostatic press with 3000 MPa. The green parts were sintered under vacuum at a temperature of 1200--1220 $^\circ$C, solution annealed at 1170--1200 $^\circ$C, and then quenched with an inert gas to room temperature. Subsequently, the parts were tempered at 850 $^\circ$C, slowly cooled to 400 $^\circ$C, and then quenched to room temperature.

The material was produced by Arnold Magnetic Technologies, and has an overall chemical composition Sm(Co$_{0.695}$,Fe$_{0.213}$,Cu$_{0.07}$,Zr$_{0.022}$)$_{7.57}$ with a minor additional oxygen content in the form of Sm$_2$O$_3$. 
\subsection{Magnetometry:}

The magnetization of a small sample piece along the easy axis was measured as a function of external field at room temperature using a Superconducting Quantum Interference Device (SQUID) in a Magnetic Property Measurement System (MPMS3) by Quantum Design. 

\subsection{Transmission Electron Microscopy:}

Electron-transparent specimens for TEM studies were prepared using Ga+ sputtering and a conventional lift-out method in Helios 600i dual-beam focused ion beam (FIB) scanning electron microscope (SEM) workstation. The ion-beam induced damage on the surfaces was reduced by low-energy ( $<$ 1~keV) Ar+ milling using a Fischione Nanomill system. The thickness of the lamellae was measured on an FEI Tecnai F30 FEG transmission electron microscope using an electron energy loss spectroscopy (EELS) log-ratio technique. A uniformly varying range of thicknesses between 80 - 140 nm was achieved.

The Sm--Co specimens were studied at remanence in magnetic-field-free (Lorentz mode) conditions using a spherical aberration-corrected FEI Titan microscope operated at 300~keV. In Fresnel-mode LTEM images, the intensity distribution at defocus $\Delta z$ is recorded to reveal a bright (convergent) or dark (divergent) contrast at the positions of the magnetic domain walls. The net deflection of electrons from the magnetic domains is induced by the Lorentz force, $\textbf{F}= -e\textbf{v}\times \textbf{B}$, where $e$ is the electron charge, $\textbf{v}$ is the velocity of the incident electrons and $\textbf{B}$ is the in-plane magnetic induction in the sample. 
A conventional microscope objective lens was used to apply magnetic fields on the specimen. TEM images were recorded using a direct-electron counting Gatan K2-IS camera and Gatan Microscopy Suite software. Electron holograms were recorded in Lorentz mode using a biprism positioned in one of the conjugated image planes of the electron column. The biprism voltage used was typically in the range of 90-100 V that forms a fringe spacing of 3 nm with a contrast of 75\%.

\subsection{Micromagnetic simulations:}

High-resolution micromagnetic simulations were performed to investigate the link between the microstructure and the domain-wall network in Sm--Co magnets. The total energy density of the system consists of: (i) ferromagnetic exchange; (ii) uniaxial magnetocrystalline anisotropy; (iii) Zeeman coupling to an external magnetic field; and (iv) dipole-dipole interactions: 

\begin{equation}
\begin{aligned}
F=& \sum_i [A^i ({{\bf \nabla \cdot m}^i})^2-K_\textrm{u}^i(m_c^i)^2 + \mu_0 M_\textrm{S}^i{{\bf m}^i \cdot {\bf H}}_{\textrm{ext}} \\
& -\frac{\mu_0 M_\mathrm{s}^i}{2}{{\bf m}^i \cdot {\bf H}}_{\textrm{dip}}^i - \mu_0 M_\textrm{S}^i{{\bf m}^i \cdot {\bf H}}_{\textrm{exc}}]\; ,
\end{aligned}
\end{equation}

\noindent where ${\bf m}^i={\bf M}^i/M_\textrm{S}^i$ is the magnetization unit vector for phase $i$ with $M_\textrm{S}^i$ the saturation magnetization, $A^i$ is the exchange stiffness, $K_\textrm{u}^i$ is the first-order uniaxial anisotropy constant, $\bf {H}_{\textrm{ext}}$ is the external magnetic field, $\bf{H}_{\textrm{dip}}$ is the local demagnetizing field due to dipole-dipole interactions, and $\bf {H}_{\textrm{exc}}$ is the exchange field at the interfaces between the different phases. The $c$-component of the magnetization is inside the lamella plane. 

The material parameters were taken from the literature \cite{katter1996,nagel2003,duerrschnabel2017}, and are $M_\mathrm{s} = 1.05$ T, $A = 23.6$ pJ/m, and $K_\mathrm{u} = 17.2$ MJ/m$^3$ for SmCo$_5$; $M_\mathrm{s} = 1.25$ T, $A = 24.7$ pJ/m, and $K_\mathrm{u} = 3.3$ MJ/m$^3$ for Sm$_2$Co$_{17}$; and $M_\mathrm{s} = 0.37$ T, $A = 11$ pJ/m, and $K_\mathrm{u} = 0$ MJ/m$^3$ for the Z phase. Note that Cu in the SmCo$_5$ changes the material parameters by lowering $M_\mathrm{s}$, $A$ and $K_\mathrm{u}$, but the composition of Cu in our material is less than 10\% of the 3d metal content and the material parameters are not strongly reduced \cite{SepehriAmin2017}. We also performed tests with the material parameters of Sm(Co$_{0.9}$Cu$_{0.1}$)$_5$ and found no qualitative difference in the behavior of the system. The simulations were tested for the lamella thickness between 50 and 100 nm, and qualitatively the same results have been obtained.

The exchange field between two phases $i$ and $j$ is proportional to $\frac{A^i}{M_\mathrm{s}^i}\frac{A^j}{M_\mathrm{s}^j}/\left({\frac{A^i}{M_\mathrm{s}^i}+\frac{A^j}{M_\mathrm{s}^j}}\right)$. Based on our optimization, described in the main text, we found the following exchange values: (i) hard to soft: 16 pJ/m; (ii) soft to hard: 13 pJ/m; (iii) soft to Z phase: 4.4 pJ/m; (iv) Z-phase to soft: 1.3 pJ/m; (v) hard to Z phase: 2.7pJ/m; and (vi) Z phase to hard: 0.95 pJ/m.

Using Equation 1, we solved the Landau-Lifshitz-Gilbert (LLG) equation of motion 

\begin{equation} \label{eq:LLG} 
\partial _{t} {\bf m}=-\gamma ({\bf m \times H}_{\textrm{eff}})+\alpha({\bf m \times} \partial _{t} {\bf m}) \; ,
\end{equation} 

\noindent where $\gamma$ is the electron gyromagnetic ratio, $\alpha$ is the dimensionless damping parameter, and ${\bf H}_{\textrm{eff}}=-{\bf \partial _{m}} F / \mu _{0} M_\textrm{S}$ is the effective magnetic field in the material consisting of external and internal magnetic fields, which depend on the material parameters. The simulations have been done with mumax3 \cite{vansteenkiste2014}, and the visualization of the magnetization textures was done with Paraview \cite{paraview}.

\subsection{Atom probe tomography:}

The needle-shaped geometry required for APT analysis was prepared by applying standard lift-out practices using an FEI Helios Focused Ion Beam 600i workstation, and mounting it to a flat-top microtip coupon supplied by Cameca. Sequential annular milling was applied to achieve an apex of $<$70 nm diameter, including low-kV cleaning, resulting in $<$0.01 at.\% Ga in the top 10 nm of the specimen. 
Data collection was performed using a LEAP4000X-HR instrument applying 100 pJ laser pulse energy with 200 kHz repetition rate and a specimen temperature of 54 K (resulting in a Co charge-state ratio (Co++/Co+) between 5 and 10). With these parameters and a chamber vacuum level at 10\textsuperscript{-9} Pa, data were collected between 5 kV and 9.5 kV with a background level consistently below 20 ppm/ns.
The atom-map reconstruction was validated by considering that the Z-platelets are atomically flat, and spatial distribution maps were performed along the c-axis (normal to the platelets) to measure the lattice spacings and thereby validate the accuracy of the atom-map reconstruction dimensions.

\subsection{Magnetic phase image and LTEM simulations:}

The electromagnetic phase shift induced in an electron wave by passing
through a sample is described by the Aharonov-Bohm effect and can
be expressed as \cite{DuninBorkowski.2004}:
\begin{align}
\varphi(x,y) & =\varphi_{\mathrm{el}}(x,y)+\varphi_{\mathrm{mag}}(x,y)\\
& =C_{\mathrm{el}}\int V(\boldsymbol{{\rm r}})dz-\frac{\pi}{\Phi_{0}}\int A_{z}(\boldsymbol{{\rm r}})dz,\label{eq:Full phase equation (both integrals)}
\end{align}
with $\varphi_{\mathrm{el}}\left(x,y\right)$ and $\varphi_{\mathrm{mag}}\left(x,y\right)$
denoting the electrostatic and magnetic contributions to the phase
shift, the interaction constant $C_{\mathrm{el}}=\frac{\gamma m_{\mathrm{el}}e\lambda}{\hbar^{2}}$,
the magnetic flux quantum $\Phi_{0}=\pi\hbar/e$, the Lorentz factor
$\gamma$, the electron rest mass $m_{\mathrm{el}}$ and the electron
wavelength $\lambda$. Furthermore, $A_{z}\left(\boldsymbol{{\rm r}}\right)$ with $\boldsymbol{{\rm r}}=\left(x,y,z\right)$
is the $z$~component of the magnetic vector potential $\boldsymbol{{\rm A}}(\boldsymbol{{\rm r}})$,
where $z$ corresponds to the incident electron beam direction \cite{Ehrenberg.1949,Aharonov.1959}.

The magnetization $\boldsymbol{{\rm M}}\left(\boldsymbol{{\rm r}}\right)$ in
the sample is linked to the vector potential by the vector convolution
integral \cite{Mansuripur.1991}
\begin{align}
\boldsymbol{{\rm A}}(\boldsymbol{{\rm r}}) & =\frac{\mu_{0}}{4\pi}\int\boldsymbol{{\rm M}}(\boldsymbol{\rm r}')\times\frac{\boldsymbol{{\rm r}}-\boldsymbol{{\rm r}}'}{\left|\boldsymbol{{\rm r}}-\boldsymbol{{\rm r}}'\right|^{3}}d\boldsymbol{{\rm r}}',\label{eq:Vector potential A(M)}
\end{align}
where $\mu_{0}$ is the vacuum permeability is the vacuum permeability. Using both equations, the magnetic phase
shift can be expressed in terms of the magnetization as

\begin{align}
\varphi_{\mathrm{mag}}\left(x,y\right) & =-\frac{\mu_{0}}{2\Phi_{0}}\int\frac{\left(y-y'\right)M_{x}(\boldsymbol{{\rm r}}')-\left(x-x'\right)M_{y}(\boldsymbol{{\rm r}}')}{\left(x-x'\right)^{2}+\left(y-y'\right)^{2}}d\boldsymbol{{\rm r}}'.\label{eq:Magnetic phase shift (arb. M)}
\end{align}
By discretizing this equation and utilizing known analytical solutions
for the magnetic phase of simple magnetized geometries, magnetic
phase images $\varphi_{\mathrm{mag}}\left(x,y\right)$ can be calculated
for arbitrary magnetization distributions $\boldsymbol{{\rm M}}\left(\boldsymbol{{\rm r}}\right)$
\cite{Caron.2018}. 

"Contour maps" are used for the visualization of the
magnetic phase in the form of magnetic induction. They are generated in Figs. 2 and 4 by taking the cosine of the magnetic phase
$\varphi_{\mathrm{mag}}$, which can be amplified beforehand to increase
the number of fringes for visualization purposes. A color scheme is
superimposed on the magnetic induction maps, which is determined by
the gradient of $\varphi_{\mathrm{mag}}$. The latter is an indicator
of the direction of the projected in-plane magnetic induction and
is indicated in Figs. 2e and 4d as a color wheel. The phase difference between two neighboring contours is $2\uppi$.

The magnetic phase $\varphi_{\mathrm{mag}}$ can further be utilized
to simulate LTEM images by convolving the corresponding wave function
$\Psi\left(x,y\right)=e^{i\varphi_{\mathrm{mag}}\left(x,y\right)}$
with a phase plate:

\begin{equation}
\Psi_{\mathrm{LTEM}}\left(x,y\right)=\mathcal{F}_{2}^{-1}\left\{ \mathcal{F}_{2}\left\{ e^{i\varphi_{\mathrm{mag}}\left(x,y\right)}\right\} \cdot e^{-i\chi\left(q_{x},q_{y}\right)}\right\} ,
\end{equation}

\noindent with $\mathcal{F}_{2}\left\{ ...\right\} $ denoting the 2D Fourier
transform, $\mathcal{F}_{2}^{-1}\left\{ ...\right\} $ its inverse
and $\chi\left(q_{x},q_{y}\right)$ denoting an aberration function
\cite{Barthel.2018} in the diffraction space containing the defocus $C_{1}$
(with positive $C_{1}$ referring to overfocus) given by:
\begin{equation}
\chi\left(q_{x},q_{y}\right)=\pi\lambda C_{1}\left(q_{x}^{2}+q_{y}^{2}\right).
\end{equation}

\noindent The LTEM images are then calculated from the corresponding electron
wave by:
\begin{equation}
I_{\mathrm{LTEM}}\left(x,y\right)=\Psi_{\mathrm{LTEM}}\left(x,y\right)\cdot\Psi_{\mathrm{LTEM}}^{*}\left(x,y\right).
\end{equation}

\section{Supplementary information}

Supplementary Figure 1 shows for the sample of Fig. 1 in the main text a diffraction pattern that contains the reflections from [100] and [110] directions. The experimental and simulated patterns reveal an excellent agreement.

\renewcommand{\figurename}{Supplementary Fig.}
\setcounter{figure}{0}

\begin{figure}[!h]
\centering
\includegraphics[width=1.0\columnwidth]{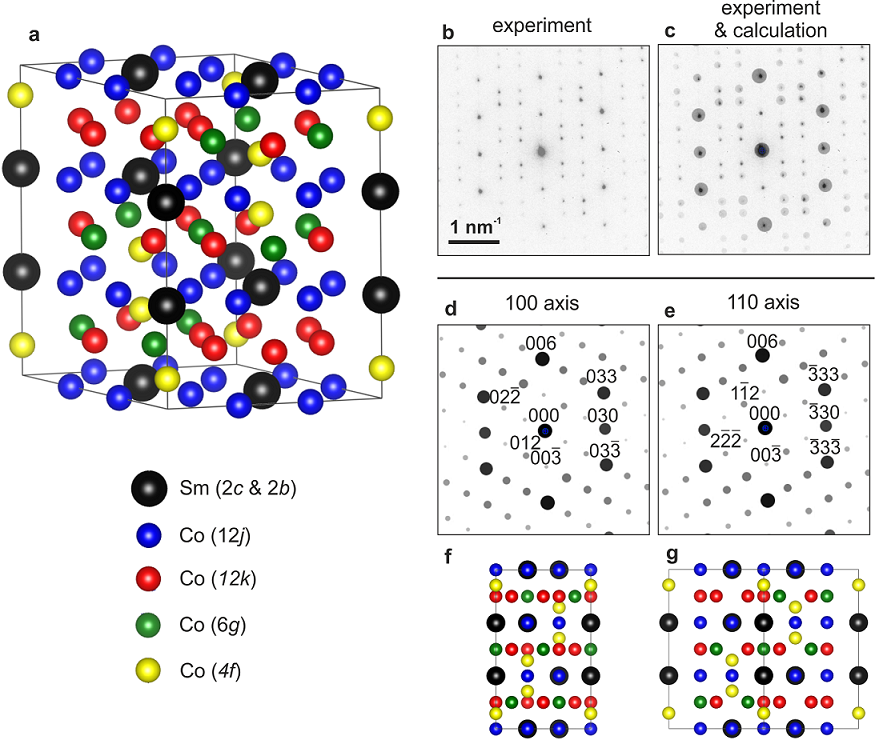}
\caption{\textbf{Crystal structure of Sm--Co:} \textbf{a} Unit cell of the Sm$_2$Co$_{17}$ phase showing the atomic sites for Sm (black) 2$b$ and 2$c$, and the 4 different atomic sites for Co: (blue) 12$j$, (red) 12$k$, (green) 6$g$, and (yellow) 4$f$.  \textbf{b} shows the recorded diffraction pattern from one Sm$_2$Co$_{17}$ cell and \textbf{c} shows the calculated pattern overlaid on the experimental pattern, illustrating an excellent agreement. The calculated pattern is a convolution of \textbf{d} [100]
 and \textbf{e} [110] reflections, and the corresponding crystal structure viewed along these directions is shown in \textbf{f} and \textbf{g}, respectively.}
\label{Diffraction}
\end{figure}

The width of the domain-wall contrast of 4 $\pm$ 2 nm at zero defocus was extrapolated from a series of LTEM images recorded at different defocus values, as shown in Supplementary Fig. 2. It is worth mentioning that this approach usually gives a slight overestimate of the true domain-wall width. 

\begin{figure}[!h]
\centering
\includegraphics[width=0.8\columnwidth]{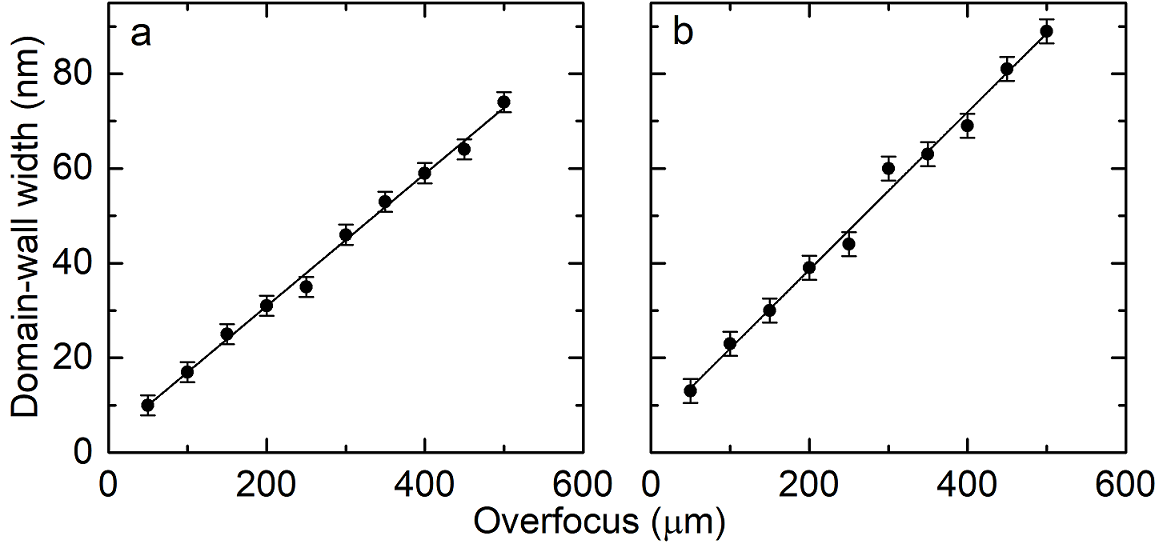}
\caption{\textbf{Measurements of domain-wall width.} \textbf{a-b} The thickness of two different domain walls was obtained from a focal-series reconstruction. The apparent width of divergent domain walls (appearing as dark contrast) was measured for different values of defocus (data points). The real width of the domain walls was deduced from the intersection of the linear fit to the data points with the y-axis.}
\label{DW width}
\end{figure}

An atom-probe tomography (APT) reconstruction is shown in the Supplementary Video 1. Isoconcentration surfaces of 9.83 at. \% Zr indicate flat Z-phase platelets, while the isoconcentration surfaces of 14.51 at. \% Sm reveal a twisted SmCo$_{5}$ cell. Note the accumulation of Cu inside the cells (pink).

\section*{Acknowledgments}
LP, MC, RES and JFL gratefully acknowledge funding from the Swiss National Science Foundation (Grant No. 200021--172934). We also thank L. Grafulha Morales and A.-G. Bitterman for the preparation of TEM samples, and ScopeM ETH Zurich for the use of its facilities.

\section*{Correspondence} 
Correspondence and requests for materials should be addressed to LP~(email: leonardo.pierobon@mat.ethz.ch), JFL~(email: joerg.loeffler@mat.ethz.ch) and MC~(email: michalis.charilaou@louisiana.edu).

\end{document}